\documentclass[prl,preprint]{revtex4}
\usepackage{epsfig}
\usepackage{amsmath}
\usepackage{epsfig}

\begin{document}
\title
{Expectation value of $p^6$ in continuous two-piece symmetric potential wells }    
\author{Zafar Ahmed$^{1,3}$ and Sachin Kumar$^2$}
\affiliation{$^1$Nuclear Physics Division, $^2$Theoretical Physics Section \\
Bhabha Atomic Research Centre, Mumbai 400 085, India\\
$~^3$Homi Bhabha National Institute, Mumbai 400 094 , India}
\date{\today}
\begin{abstract}
Earlier, potentials like square well and several other half-potential wells with discontinuous jump have been found to have the expectation value  $<\! p^6 \!>$ to be divergent for all bound states. Here, we consider two-piece symmetric potential wells to prove and demonstrate that in them the expectation value of $p^6$ diverges for even states and converges for odd states. Here, $p$ denotes momentum. We also present three exactly solvable models. 
\end{abstract}
\maketitle
In quantum mechanics [1] for a potential well the expectation value $<\!\psi(x)|F|\psi(x)\!>$ of an operator $F$ is obtained using  eigenfunction of a bound state that  is a continuous and normalizable solution of Schr{\"o}dinger equation
\begin{equation}
\frac{d^2\psi(x)}{dx^2}+\frac{2m}{\hbar^2}[E-V(x)] \psi(x)=0.
\end{equation}
It is also be desirable that uncertainty in position $\Delta x$ and in momentum $\Delta p$ should be finite to meet the Heisenberg's uncertainty principle [1].
Recently, it has been remarked that for $\Delta x$ to be finite the  eigenstate in position space needs to vanish faster than $1/|x|^{3/2}$ [2]. For instance, for $\psi_0(x)=\frac{1}{\sqrt{1+x^2}}$, $\Delta x = \infty$ and the potential possessing it 
is a double humped well (a well with )with two side barriers) [2], where $\psi_)(x)$ is the ground state at the $E=0$. Other wise all the usual potential wells like Dirac delta well, square well, harmonic well and other well potentials have finite value for $\Delta x$.

Students are also advised to work in momentum representation [1] where the wave function $\phi(p)$ is given as
\begin{equation}
\phi(p)={\cal F}[\psi(x)]=(2\pi \hbar)^{-1/2} \int_{-\infty}^{\infty} \psi(x) e^{-ipx/\hbar} dx,
\end{equation}
the Fourier transform of  $\psi(x)$: ${\cal F}$$[\psi(x)]$. The two representations are physically equivalent.
One can find $<\!x^2\!>$ as $<\!\psi(x)|x^2|\psi(x)\!>$ or $<\!\phi(p)|-\hbar^2(d^2/dp^2)|\phi(p)\!>$. Similarly, $<p^2>$ can be found
as $<\!\psi(x)|-\hbar^2(d^2/dx^2)|\psi(x)\!>$ or $<\!\phi(p)|p^2|\phi(p)\!>$. 
Instead of finding the Fourier transform (2) of $\psi(x)$, one can in principle solve the integral equation
\begin{equation}
[p^2/2m -E] \phi(p)= \int_{-\infty}^{\infty} U(p-p') \phi(p') dp'.
\end{equation}
directly for $\phi(p)$, where $U(p)$=${\cal F}[V(x)]$. Solving the integral equation (3) may be more difficult. Mostly the mathematical forms of $\psi(x)$
and $\phi(p)$ are different so much so that for finding something, one option is either easier to do or more transparent than the other one.
Also these two options present different mathematical situations. For instance, for infinite square well (ISW),  the $p$-integral
in finding $<p^2>$ is improper [3] but finite whereas the $x$-integrals are proper and simple. Once again, in order to have $\Delta p$ to be finite, we may demand that $\phi(p)$ needs to fall off faster than $1/|p|^{3/2}$. One may find $<p^2>$ more transparently and directly as
\begin{equation}
<\!\psi_n|p^2|\psi_n\!>  =\frac{2m}{\hbar^2} \int_{-\infty}^{\infty} \psi_n(x) [E_n-V(x)]~ \psi_n(x) dx,
\end{equation}
If the potential well is finite, the bound state eigenfunctions will be finite, continuous and differentiable for each $x \in (-\infty, \infty)$, so will be the expectation value of $p^2$. Even for the Dirac Delta well (3) is finite.

For ISW, $<\!p^4\!>$ in the position representation gives a finite
value, it actually diverges in momentum space. Similar experience is found  in finite square well (FSW) where it is $<\!p^6\!>$ which presents an interesting  discrepancy in the two representations.This discrepancy was first pointed out in a largely un-noticed paper [4] where for FSW $I_n(p)=|\phi_n(p)|^2$ was derived to show a surprising asymptotic fall-off as $p^{-6}$, however the details of $\phi_n(p)$ were incorrect which have been corrected recently [5]. The consequent  divergence of $<\! p^6 \!>$ in FSW in position space  was revealed in terms of the Dirac delta discontinuities in the second and higher order derivatives of $\psi_n(x)$ at the end points $x=\pm a$. Unfortunately, this proof is not very transparent [4].  Recently, a  simple and transparent proof for the divergence of $<\!p^6\!>$ for square well potential has been presented [5]. Following this, it has been shown that $<\!p^6\!>$ diverges when potential wells have a finite jump discontinuity. These potentials [6] are two-piece half-potential wells of the type $V(x)=-U(x) \Theta(x)$, $\Theta(x)$ is Heaviside step function defined as $\Theta(x<0)=0, \Theta(x>0)=1$ and $U(x)$ is a differentiable function which may or may not vanish asymptotically and $U(0)=-V_0$.

Here, in this paper we wish to state, prove and demonstrate that of $<\!p^6\!>$ diverges (converges)
for even (odd) parity states of continuous two-piece symmetric potential wells which are non-differentiable at $x=0$. These potential wells  have left and right derivatives mismatching at $x=0$, these are function of $|x|$ like $V_1(x)=-V_0 |x|/a$ [7], $V_2= V_0 \exp(-2|x|/a)$ [8] and $V(x)=V_0[\exp(2|x|/a)-1]$ [9]. Such potential wells enrich students' experience in studying bound states in quantum mechanics.
\begin{figure}
	\centering
	\includegraphics[width=4.5 cm,height=3 cm]{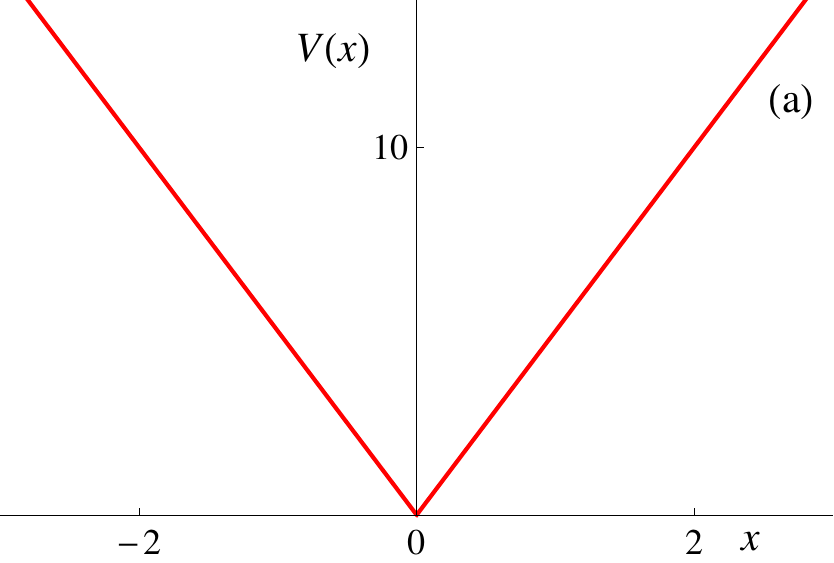} 
	\hskip .1 cm
		\includegraphics[width=4.5 cm,height=3 cm]{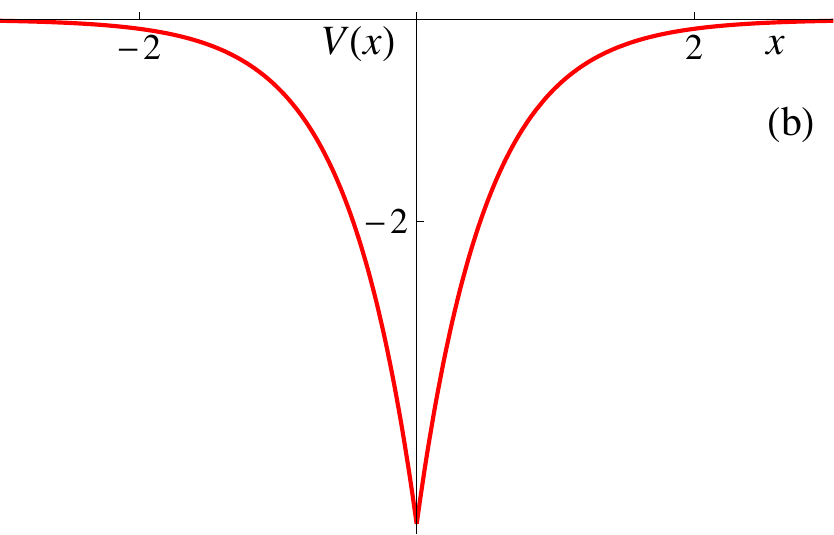} 
		\hskip .1 cm
		\includegraphics[width=4.5 cm,height=3 cm]{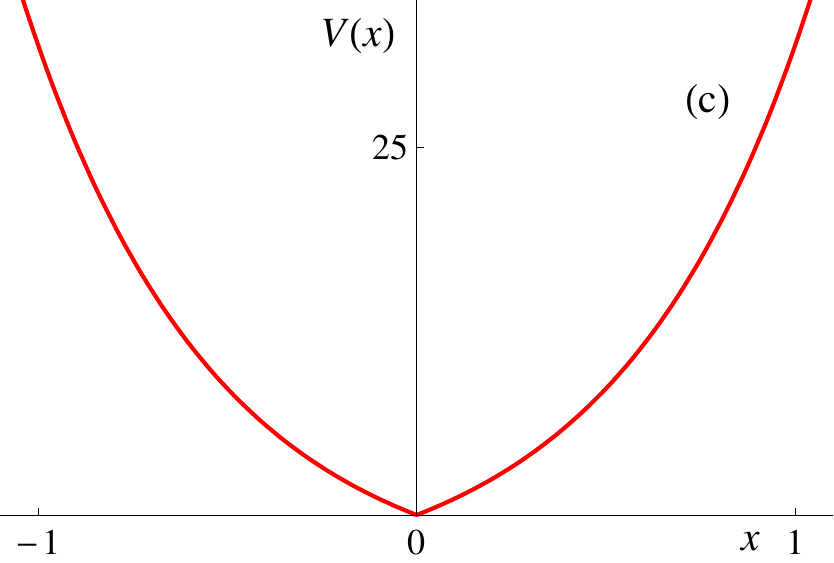} 
	\caption{Schematic representation of continuous two-piece  symmetric potential wells  which are non-differentiable at $x=0$. (a): triangular well, $V_1(x)=-V_0 |x|/a$ (14), (b): convergent exponential well, $V_2(x)=-V_0\exp(-2|x|/a)$ (18) and (c): divergent exponential well, $V_3(x)=V_0[\exp(2|x|/a)-1]$ (22).}
\end{figure}

A  definite integral $\int_{a}^{b} f(x) dx$ is real and finite if it is continuous at each and every point of the domain $[a,b]$, $f(x)$ may also be piece-wise continuous for this integral to exist. Otherwise, the integrals are improper which may be convergent (finite) or divergent (infinite) [2].  $V(x)=-2\delta(x)$ is an  interesting digression 
where in $<\!p^2\!>$ is finite owing to the interesting property that $\int_{-\infty}^{\infty} f(x) \delta(x) dx =f(0).$ 

Next, we suggest $<\! p^4 \!>$ to be evaluated  as 
\begin{equation}
- \int_{-\infty}^{\infty} \psi(x) \frac{d^2}{dx^2}[E-V(x)] \psi(x) dx,
\end{equation}
which can be re-written in an inspiring form as
\begin{equation}
<\!p^4\!{>}{=}{\int_{-\infty}^{\infty}}{F_1[\psi,\psi',{V'},V{''}]} dx{+}{<}\psi{|(E{-}V(x))^2|}\psi{>}. 
\end{equation}

Ordinarily, the first integral in above simplifies to $[-V'(x) \psi^2(x)]_{-\infty}^{\infty}$. When $V(x)$ is continuous and differentiable, it vanishes since $\psi(x)$ are bound states
that converge to zero, asymptotically. Alternatively,  inside the first integral in (6),  there occur  terms like $2[V'(x) \psi(x) \psi'(x)+V''(x) \psi^2(x)]$. For the Dirac delta well $V(x)=-2\delta(x)$ using the interesting derivatives   $V'(x)=2\delta(x)/x$ and $V''=-4\delta(x)/x^2$; $\psi_0(x)=e^{-|x|}$, the second term causes strong divergence in $<\!p^4\!>$ near $x=0$ as
\begin{equation}
\int_{-\epsilon}^{\epsilon} V''(x) \psi^2(x) dx=-4 \int_{-\epsilon}^{\epsilon} e^{-2|x|} \frac{\delta(x)}{x^2} dx \rightarrow \infty.
\end{equation}
Had there been odd eigenstate(s) this integral would have been convergent and finite. This explains the divergence of $<\! p^4 \!>$ in position space which is obvious in momentum space as $\phi(p)=\sqrt{2/\pi} (1+p^2)^{-1} [2].$ 
Next, we verify that expectation value of force $(-V'(x))$, namely
\begin{equation}
<\!\psi{|V'(x)|}\psi\!{>}= 2\int_{{-}\epsilon}^{\epsilon}\frac{\delta(x)}{x}\psi^2(x)dx{\rightarrow} 2\int_{{-}\epsilon}^{\epsilon} \frac{\delta(x)}{x}(1{-}2{|}x{|}) dx 
\end{equation}
vanishes as here is an odd integrand between symmetric limits. Vanishing of integrals in (8), may not be without arguments. Here, we underline that otherwise Ehrenfest theorem will be defied by
the Dirac delta well potential which is most popular among potential wells. 

{\bf Convergence of $<\! p^4 \!>$ for $V(x)=|x|, e^{-2|x|}, e^{2|x|}-1$  Fig. 1} 

Near, $x=0$, the even parity bound states of a symmetric well near $x=0$ behave as $\psi_e(x) \approx A$ and the odd ones behave as $\psi_o(x) \approx B x$. For the potential well $V_1(x)=  |x|$, we have $V_1'(x)= S(x)$, where $S(x<0)=-1$ and $S(x>0)=1$ such that $S(x)=2 \Theta(x)-1$, $\Theta'(x)=\delta(x)$, then $V_1''(x)= 2\delta(x)$, $V_1'''(x)=2\delta'(x)$ and $V_1^{iv}(x)=  2 \delta ''(x)$. The $n^{th}$ derivative of $\delta(x)$ namely $\delta^{(n)}(x)=(-1)^n n! ~\delta(x)/x^n$  So we find that in the $\epsilon$-vicinity of $x=0$ the crucial part of $<\! p^4 \!>$ behaves as
\begin{equation}
<\psi|V_1''|\psi> = 2 \int_{-\epsilon}^{\epsilon} \psi^2(x) \delta(x) dx \rightarrow \mbox{finite},
\end{equation}
Next, for $V_2(x)=\exp(-2|x|)$, $V_2'(x)=-2e^{-2|x|} S(x)$, $V_2''(x)=4 \exp(-2|x|) S^2(x)-4\exp(-2|x|) \delta(x)$. So  for $V_2(x)$
\begin{equation}
<\psi|V_2''|\psi> = 4\int_{-\epsilon}^{\epsilon} \exp(-2|x|)~~[1-\delta(x)] \psi^2(x)
~dx
\end{equation}
remains finite and hence $<\! p^4 \!>$. Similar steps justify the convergence of $<\!p^4\!>$
for $V_3(x)$ as well.

{\bf Expectation value of $p^6$ for three potentials in Fig. 1:}\\
Using $p^2 \psi(x)=[(E-V(x)]\psi(x)$ successively and Eq. (6) wet
\begin{equation}
<\!p^6\!>{=}\int_{{-}\infty}^{\infty} {F_2[}\psi,\psi',V,{V'},{V''},{V'''},V^{(iv)}] dx\\{+}<\!\psi|[E-V(x)]^3|\psi \!>.
\end{equation}

For the expectation value of $<\!p^6\!>$, we get 
In the above equation the part $<\! \psi(x)|V^{iv}(x)| \psi(x)\!>$ is the main source of divergence in $<\!p^6\!>$.

For $V_1(x)=|x|$, we have $V^{iv}_1(x)=4 \delta(x)/x^2$
\begin{equation}
<V^{iv}_1(x)> = 4 \int_{-\epsilon}^{\epsilon} \frac{\delta(x)}{x^2} \psi^2(x) dx,
\end{equation}
which diverges for the even state ($\psi_e(x) \approx A$) and converges for odd state. 
For $V_{2}(x)= \exp(-2|x|)$, by successive integrations,  we find that  the strongest divergent  term in $V^{iv}_2(x)$ is $-2 \exp(-2|x|)
~\delta^{'''}(x)$
\begin{equation}
<\!V^{iv}_2(x)\!> = 12 \int_{-\epsilon}^{\epsilon} \exp(-2|x|)  \frac{\delta(x)}{x^3} \psi^2(x) dx,
\end{equation}
which diverges for even parity state $(\psi_e(x) \approx A)$ and converges to zero for the odd one $(\psi_o(x) \approx B x)$ as the integrand has  odd parity. Similar results follow for $V_3(x)$. The expectation of higher even powers of $p$, e.g., $<\! p^8\!>$ will consist of $<\! V^{vi}(x)\!>$, this in turn will involve $\delta^{iv}(x)$ and hence it will diverge.

In the following, we study the momentum distributions $p^2 I_n(p), p^4 I_n(p)$ and $p^6 I_n(p)$ for
three exactly solvable models for the ground state  and  the first excited  state by finding  $\phi_n(p)$ from their position space eigenfunction $\psi_n(x)$ using the Fourier transform (2).
The momentum distribution $I_n(p)$ is obtained as $|\phi_n(p)|^2$.
\vskip .2 cm
{\bf 1. Symmetric triangular well:} This potential is given as
\begin{equation}
V(x)=V_0 \frac{|x|}{a}, V_0>0.
\end{equation}
The Schr{\"o}dinger equation (1) for this potential (14)   can be transformed to the Airy differential equation [7,10] as
\begin{equation}
\frac{d^2\psi}{dy^2}{-}y \psi{=}0, \quad y(x){=}\frac{2m}{g^2\hbar^2}\left[\frac{V_0|x|}{a}{-}E\right],\quad g{=}\sqrt[3]{\frac{2mV_0}{\hbar^2 a}}.
\end{equation}
This second order equation has two linearly independent solutions called Airy functions $Ai(y)$ and $Bi(y)$. It is $Ai(y)$ that vanishes as $x \sim \infty$, so we admit  the solution of  (14)
as $\psi(x\ge 0) = C Ai(y(x))$ and for $x<0$, we have $\psi(x < 0)= C Ai(y(-x))$. For the even parity state, we demand $\psi'(0)=0$
\begin{equation}
Ai'(y_0)=0, \quad  \psi(x)=C Ai(y(x)), \quad  y_0=-\frac{2m}{\hbar^2}\frac{E}{g^2}. 
\end{equation}
For the odd parity states we demand $\psi(0)=0$, so the eigenvalue condition and the eigen functions are
\begin{equation}
Ai(y_0)=0, \quad \psi(x)=C~ \mbox{sgn}(x) Ai(y(x)), 
\end{equation}
We take $V_0=5$ and $a=1$ in arbitrary units, the well has two bound states at  $E=2.9789$ and $E=6.8366$ as per Eqs. (16) and (17). The three momentum distributions are plotted for the first even and the first odd state in Fig. 2(a) and 2(b), respectively. $p^2 I_n(p)$ and $p^4 I_n(p)$ show fast convergence  to zero in both parts (a) and (b) but $p^6I_0(p)$ has long tail in Fig. 2(a) displaying a divergence for $<\! p^6\!>$, whereas the odd parity states presents short range characteristic of the distribution $p^6 I_1(p)$.
\begin{figure}
	\centering
	\includegraphics[width=7 cm,height=5 cm]{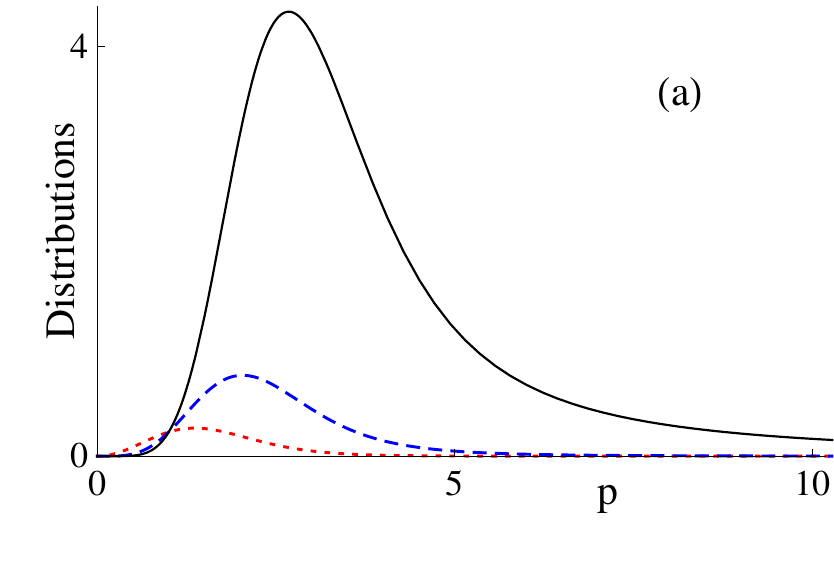}
	\hskip .1 cm
	\includegraphics[width=7 cm,height=5 cm]{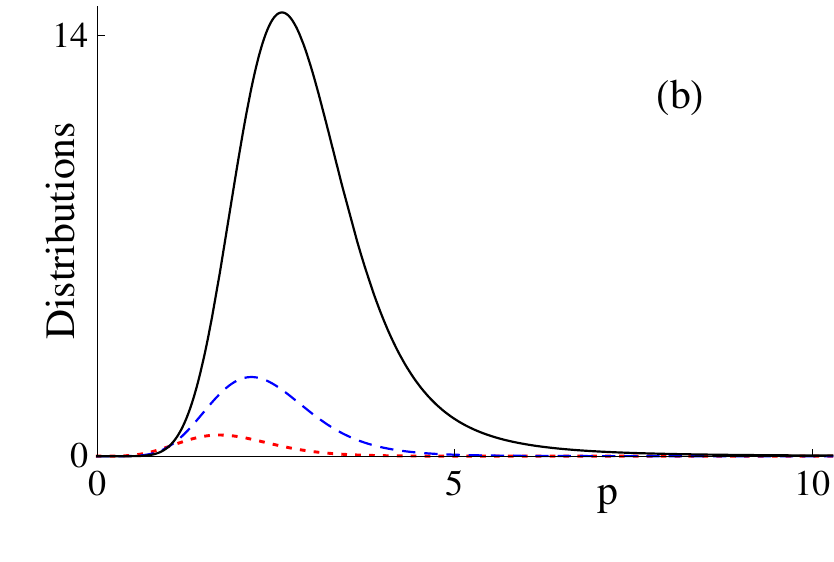}
	\caption{Various momentum distributions $p^{2j} I_n(p), j=1,2,3$ for the triangular potential well (14); (a): for first even parity state ($n=0$), (b): for the the first odd parity state ($n=1$). Distributions for $j=1,2$ are convergent, however for $j=3$ in (a) the distribution show a long tail for even parity state. }
\end{figure}
\vskip .2 cm
{\bf 2. Symmetric (convergent) exponential well:}
This potential is given as 
\begin{equation}
V(x)=-V_0 \exp(-2|x|/a), V_0>0
\end{equation}
The Schr{\"o}dinger equation for this potential can be transformed to the Bessel equation [8,10] as
\begin{eqnarray}
w^2\frac{d^2\psi}{dw^2}+ w\frac{d\psi}{dw}+(k^2 a^2+w^2) \psi=0,~ w=qae^{-|x|/a},  k=\frac{\sqrt{2m(-E)}}{\hbar}, E<0,  q=\frac{\sqrt{2mV_0}}{\hbar},
\end{eqnarray}
whose two linearly independent solutions are $J_{ka}(w)$ and $J_{-ka}(w)$. Noting that when $w$ is very small $J_{ka}(w) \approx \frac{w^{ka}}{\Gamma(1+ka)}$ so when $x>0$, $J_{ka}(w) \sim e^{-kx}$ represents bound solution. So we choose $\psi(x)= C J_{ka}(qa e^{-|x|/a})$. For even parity states, we demand $\psi'(0)=0$ to get quantization condition and the corresponding eigenfunctions as
\begin{equation}
J'_{ka}(qa)=0, \quad \psi(x)= C J_{ka}(qa e^{-|x|/a}).
\end{equation}
For the odd parity states, we demand $\psi(0)=0$ and get the quantization condition and the corresponding eigenfunctions as
\begin{equation}
J_{ka}(qa)=0, \quad  \psi(x)= C ~ \mbox{sgn}(x) J_{ka}(qa e^{-|x|/a}).
\end{equation}
For $V_0=15, a=1$ Eq. (20) yields the ground state eigenvalue as $E=-7.3460$ and Eq. (21) yields the eigenvalue of the first excited state as $E=-1.0622$.  For the first two states, we plot various distributions as in Fig. 3. One can visualize the long tail in $p^6 I_0(p)$ in Fig. 3(a) that would give rise to divergence in $<\! p^6 \!>$ for the even parity state.

\begin{figure}
	\centering
	\includegraphics[width=7 cm,height=5 cm]{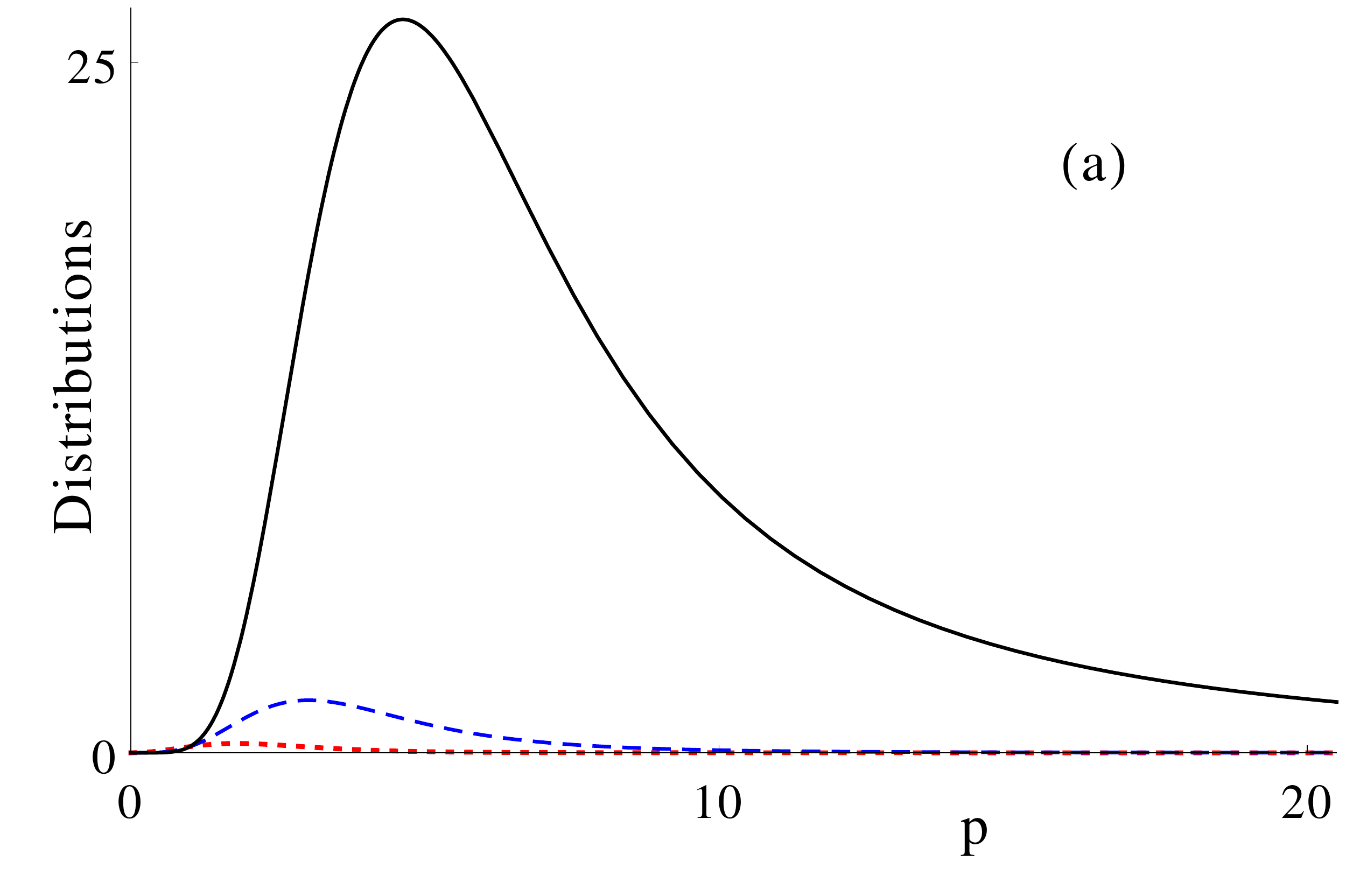}
	\hskip .1 cm
	\includegraphics[width=7 cm,height=5 cm]{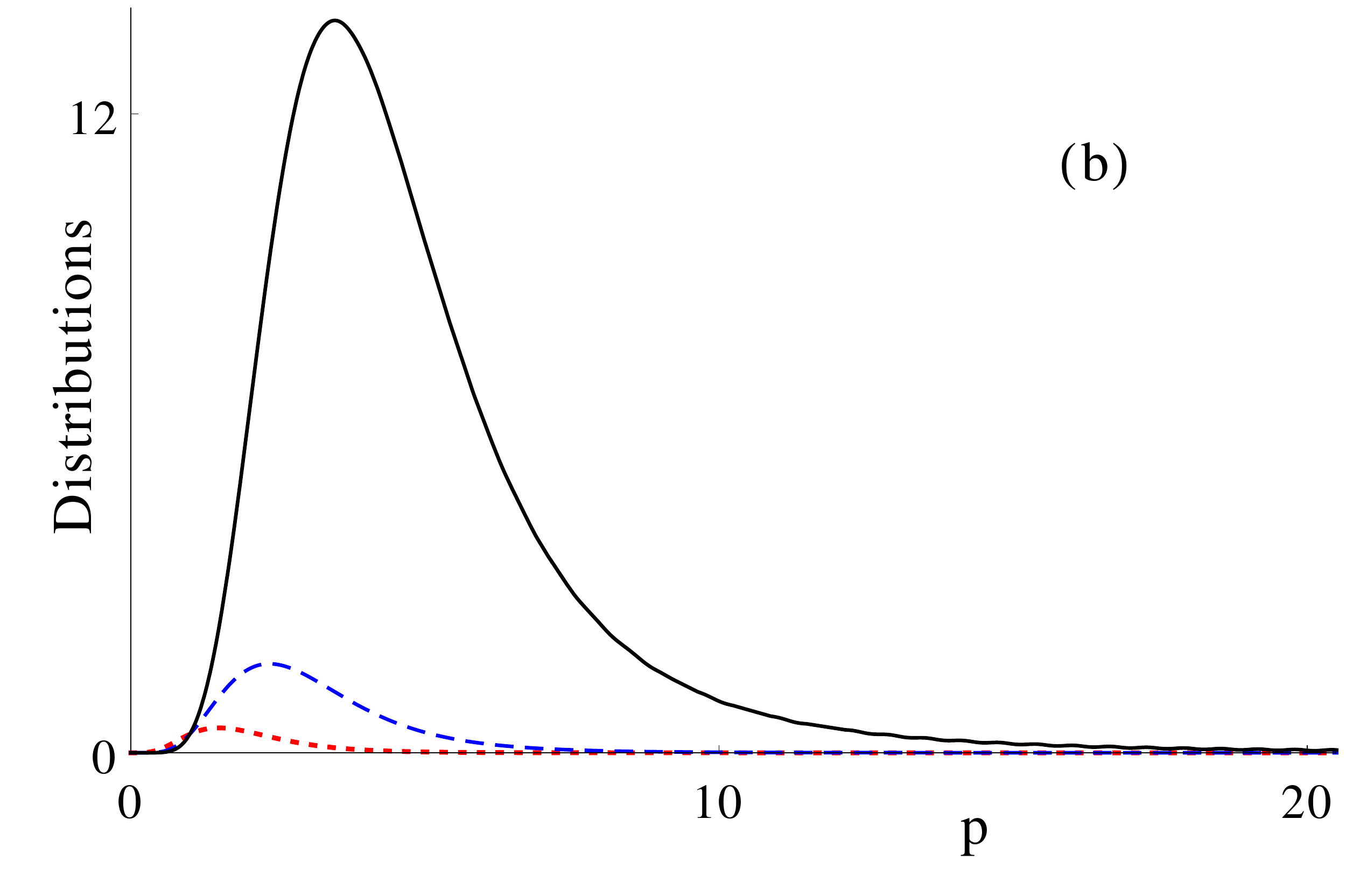}
	\caption{The same as in Fig. 2, for the convergent exponential potential well (18). Visualize the  much longer tail in (a) in the solid curve indicating the divergence of $<\! p^6 \!>$ for the even parity state.}
\end{figure}
\vskip .2 cm
{\bf 3. Symmetric (divergent) exponential well:} This potential is written as
\begin{equation}
V(x)=V_0[e^{2|x|/a}-1],
\end{equation}
for which the Schr{\"o}dinger equation (1) can be transformed to the cylindrical Bessel equation
as [9,10]
\begin{eqnarray}
z^2\frac{d^2\psi}{dz^2}+ z\frac{d\psi}{dz}+(-\kappa^2 a^2-z^2) \psi=0,\quad  z=qae^{|x|/a},  \kappa=\frac{\sqrt{2m(E+V_0)}}{\hbar},  q=\frac{\sqrt{2mV_0}}{\hbar}.
\end{eqnarray}
Out of two linearly independent solutions of (23) as modified Bessel function: $I_{i\kappa a}(z)$ and $K_{i\kappa a}(z)$. Here, we choose $K_{i\kappa a}(z)$ as the solution of (22) since it vanishes for $|x|\sim \infty$. For even parity states we demand $\psi'(0)=0$, then the quantization condition and eigenfunctions are given as
\begin{equation}
K'_{i\kappa a}(qa)=0, \quad \psi(x)=C K_{i\kappa a} (qa e^{|x|/a}).
\end{equation} 
For odd parity state we demand $\psi(0)=0$, we get the eigenvalue equation and eigenfunctions as
\begin{equation}
K_{i\kappa a}(qa)=0, \quad \psi(x)= C~ \mbox{sgn}(x) K_{i\kappa a}(qa e^{|x|/a}) .
\end{equation}
For $V_0=5$ and $a=1$,  we get first two bound states in the potential at $E=6.4646$ and $E=17.5365$. The three distributions are plotted in Fig. 4, where the solid line in Fig. 4(a) yet again indicates much longer tail justifying the divergence of $<\! p^6 \!>$ for the even parity state.
\begin{figure}
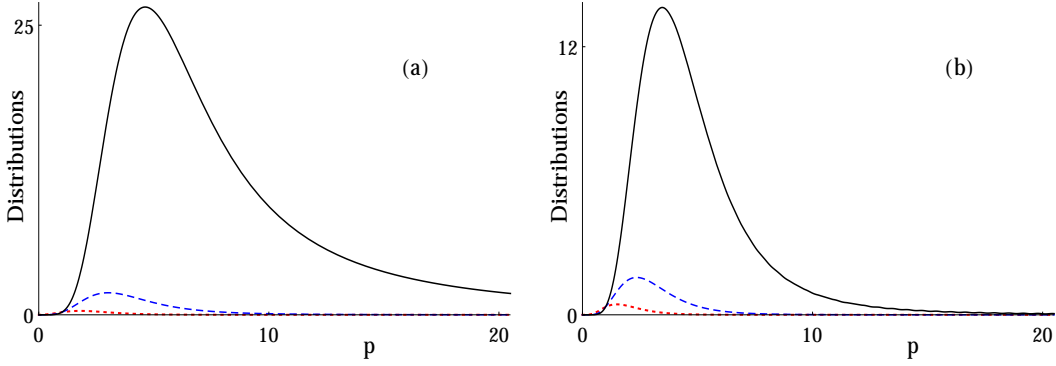

	\centering
	\includegraphics[width=7 cm,height=5 cm]{Fig-3a.pdf}
	\hskip .1 cm
	\includegraphics[width=7 cm,height=5 cm]{Fig-3b.pdf}
	\caption{The same as in Fig. 2, for the divergent exponential potential well (22). Visualize the  much longer tail in (a) in the solid curve indicating the divergence of $<\! p^6 \!>$ for the even parity state.}
\end{figure}

We would like to mention that if
$<\! p^6 \!>$ is divergent and hence $<\! p^{2j}\!>, j=4,5,6..,$ also diverge.  Also, $<\! p^{2j+1} \!>$ for $j=0,1,2,..$ vanish due to antisymmetry of the integrands. Momentum distributions for other interesting one-dimensional potential wells can be seen in Refs. [6,11].
 
 The divergence of $<\!p^6\!>$ or the long tail of $p^6 I_0(p)$ for these symmetric two piece wells though proved in Eqs. (12) and (13) in a simple and transparent way show very well in Figs. 3(a) and 4(a).  For the triangular well, the long tail may  be missed out in Fig. 2(a), however it exists there.

In one dimension, functions  which are discontinuous or non-differentiable at some point(s)
are called non-analytic. Both analytic (e.g. $V(x)=x^2, \mbox{sech}^2x$ and non-analytic  potentials (e.g., $-\delta(x)$, square well [4,5], the wells in Ref. [6] or the ones in Fig. 1 discussed here) have $<\!p^2\!>$ as finite, but in the cases of non-analytic ones, $<\! p^{2j}\!>, j=2,3,4$ may diverge as the function of $p$.
 
We hope that
the examples presented here will enrich students' experience wherein $<\! p^{6}\!>$,  diverges (converges) for even (odd) parity states of a symmetric continuous  two piece potential wells (Fig. 1) while calculating them in momentum space. These divergences which occur in non-analytic potential wells have been  demonstrated to arise in the position space due to the occurrence of higher order derivatives of Dirac delta function at the critical point $(x=0)$  of the potential well where it is non-analytic. Interestingly, otherwise the potential wells discussed here may be thought to be similar to other one piece analytic potential well such as harmonic oscillator well.

\end{document}